\documentclass[12pt,showpacs,aps,prb,amsmath,amssymb,superscriptaddress]{revtex4}
\usepackage{graphicx}
\usepackage{dcolumn}
\usepackage{bm}

\begin{document}

\baselineskip=0.8cm
\renewcommand{\thefigure}{\arabic{figure}}
\def\be{\begin{equation}}
\def\ee{\end{equation}}
\def\bea{\begin{eqnarray}}
\def\eea{\end{eqnarray}}
\def\E{{\rm e}}
\def\bearst{\begin{eqnarray*}}
\def\eearst{\end{eqnarray*}}
\def\peleven{\parbox{11cm}}
\def\peffec{\peight{\bearst\eearst}\hfill\peleven}
\def\pspace{\peight{\bearst\eearst}\hfill}
\def\ptwelve{\parbox{12cm}}
\def\peight{\parbox{8mm}}

\title{  Roughness of undoped graphene and its short-range induced gauge
field  }
\author{N. Abedpour}
\affiliation{Department of Physics, Sharif University of Technology,
P.O. Box 11365-9161, Tehran 11365, Iran}
\author{M. Neek-Amal}
\affiliation{Department of Nano-Science, IPM, Tehran 19395-5531,
Iran}
\author{Reza Asgari}
\affiliation{Institute for Studies in Theoretical Physics and
Mathematics, Tehran 19395-5531, Iran}
\author{F. Shahbazi}
\affiliation{Department of Physics, Isfahan University of
Technology, Isfahan 84156-83111, Iran}
\author{ N. Nafari}
\affiliation{Institute for Studies in Theoretical Physics and
Mathematics, Tehran 19395-5531, Iran}
\author{M. Reza Rahimi Tabar $^{1,}$  }
\affiliation{ CNRS UMR 6529, Observatoire de la C$\hat o$te d'Azur,
BP 4229, 06304 Nice Cedex 4, France \\$^5$ Carl von Ossietzky
University, Institute of Physics, D-26111 Oldenburg, Germany }

\begin{abstract}
We present both numerical and analytical study of graphene roughness
with a crystal structure including $500 \times 500$ atoms. The
roughness can effectively result in a random gauge field and has
important consequences for its electronic structure. Our results
show that its height fluctuations in small scales have scaling
behavior with a temperature dependent roughness exponent in the
interval of $ 0.6  < \chi < 0.7 $. The correlation function of
height fluctuations depends upon temperature with characteristic
length scale of $ \approx 90 {\AA}$ (at room temperature). We show
that the correlation function of the induced gauge field has a
short-range nature with correlation length of about $\simeq 2-3
{\AA}$. We also treat the problem analytically by using the
Martin-Siggia-Rose method. The renormalization group flows did not
yield any delocalized-localized transition arising from the graphene
roughness. Our results are in good agreement with recent
experimental observations.
\end{abstract}
 \maketitle
\hspace{.3in}
\section{Introduction}

An isolated two dimensional (2D) sheet of carbon atoms having a
simple honeycomb structure is known as graphene~\cite{wallace}.
According to the recent detailed measurements, these 2D zero-gap
semiconductors reveal new features in their electronic properties.
In particular, the low-energy quasi-particles of the system can
formally be described by the massless Dirac-like
fermions~\cite{novoselov}. Doped graphene sheets are pseudo-chiral
2D Fermi liquids with abnormal effective electron-electron
interaction physics~\cite{yafis}. From the experimental point of
view, the melting temperature of thin films decreases with
decreasing thickness and become unstable when their thickness
reduces to a few atomic layers~\cite{evens}. This observation
supports the theoretical prediction that states: No strictly 2D
crystal can be thermodynamically stable at finite temperatures
~\cite{mermin}. Consequently, the expectation of not observing a
free 2D material in nature lived on till the Geim's group discovered
graphene~\cite{mayer}. Furthermore, because of the existing coupling
between bending and stretching energy modes in any 2D material, one
expects to observe a measurable undulations or at least very small
roughness on the graphene sheet and will reduce electronic transport
in graphene and affecting its thermal conductivity~\cite{novoselov,
mayer}.

In this paper, we are interested in determining the temperature
dependence of graphene {\it roughness}. Here roughness is defined as
the variance of the height fluctuations in graphene due to
temperature. Our numerical treatments suggest that its height
fluctuations have a scaling behavior in small scale with a
temperature dependent roughness exponent. We also determine the
temperature dependence of the amplitudes of the height structure
function. We show that the induced random gauge field has a short-
range correlation function with correlation length $\simeq 2-3
{\AA}$. Finally, we formulate a field theoretical method to
investigate the electron dynamics of the undoped graphene in such a
random gauge field and show that there is no disorder-induced
transition from delocalized to localized states$^7$. This important
result is in good agreement with the experimental observation that
due to the graphene roughness localization is
suppressed~\cite{morozov}.

The rest of this paper is organized as follows. In Sec.\,II, we
introduce an effective interaction between carbon atoms that enters
the molecular dynamic simulation to model the roughness of graphene
. Section III contains our numerical calculations of graphene
simulation and then using the latest results calculating the $\beta$
function analytically by using the field theory of
Martin-Siggia-Rose method. We finally conclude in Sec.\,IV with a
brief summary.

\section{Roughness Exponent}

To study and arrive at a quantitative information of a graphene
surface one may consider a surface with size $L$ and define the mean
height of the surface, $\overline{h}(L,\lambda)$, and its roughness,
$w(L,t,\lambda)$, by the expressions like
$\overline{h}(L,\lambda)=\frac{1}{L}\int_{-\frac{L}{2}}^{\frac{L}{2}}h(\bf
{x},\lambda)dx $ and
$w(L,t,\lambda)=(<(h-\overline{h})^2>)^{(\frac{1}{2})}$,
respectively. The symbol $<...>$ denotes an ensemble averaging.
Here, $\lambda$ is an external factor which could be temperature in
this problem ~\cite{sang} and $t$ is the time. In the limit of large
$t$, the roughness saturates and behaves as $w(L,\lambda) \sim
L^{\chi (\lambda)}$. The roughness exponent, $\chi$, characterizes
the self-affine geometry of the surface. The common procedure for
measuring the roughness exponents of a rough surface is to use a
surface structure function, $S(r)=<|h(x+r)-h(x)|^2>$ which depends
on the length scale $\Delta x = r$. The surface structure function
is equivalent to the statistics of height-height correlation
function, $C(r)$, and are related by $S(r)=2w^2(1-C(r))$ for
stationary surfaces. The second order structure function $S(r)$
scales with $r$ as $r^{\xi_2}$ where $\chi=\xi_2/2$
~\cite{bara,jafari}.

The atomic structure of graphene will force us to define two scaling
exponents in $x-$ (zig-zag) and $y-$ (arm-chairs) directions. The
exponents can be found via the second order structure functions, $
S_x(r_x)=<|h(x+r_x,y)-h(x,y)|^2> $ and $
S_y(r_y)=<|h(x,y+r_y)-h(x,y)|^2>$. The different scaling exponents
in the zigzag and arm-chairs directions show the anisotropic nature
of the roughness in graphene. We have used the empirical
inter-atomic interaction potential, i.e., carbon-carbon interaction
in graphite~\cite{brenner}, which has in addition three-body
interaction for molecular dynamics simulation of graphene sheet to
investigate its morphology and its dependence of roughness exponents
on temperature. The two-body potential gives a description of the
formation of a chemical bond between two atoms. Moreover, the
three-body potential favors structures in which the angle between
two bonds is made by the same atom. Many-body effects of electron
system, in average,  is considered in the Brenner potential, through
the bond-order and furthermore, the potential depends on the local
environment.

It is well known that the harmonic approximation resulting in
bending instabilities due to soft wavelength phonons leads to
crumpling of a membrane. It is important to note that the Brenner
potential has anharmonic coupling between bending and stretching
modes which prevents crumpling. We have considered a graphene sheet
including a size of $500 \times 500$ atoms with periodic boundary
condition. Considering the canonical ensemble (NVT), we have
employed Nos\'{e}-Hoover thermostat to control temperature. Our
simulation time step is $1~fs$ in all cases and the thermostat's
parameter is $5~fs$. Therefore, we have found a stable 2D graphene
sheet in our simulation.

\section{Numerical results}

In the top graph of Fig.~1, we have shown a snapshot of the graphene
fluctuations at temperature $300 K$. Moreover, the magnetic field
induced by the roughness is shown in the bottom graph to emphasize
the randomness of the surface structure. Order of height
fluctuations are about $\sim 5 {\AA}$ which is in good agreement
with experimental observation\cite{mayer}. In Fig.~2, we have
plotted (in log-log scales), the structure function in the
arm-chairs direction, namely $S_y(r_y)$, simulated at temperatures
$10,30,100$ and $300$ $K$. Fig.~2 shows that for some characteristic
length scales, the scaling behavior of the structure functions do
not exist. The typical characteristic length scales are
approximately $90{\AA}$ at room temperature which is in good
agreement with experimental findings~\cite{mayer} that is in the
range of $50-100{\AA}$.

In Fig.~3, the temperature dependence of the scaling exponents,
$\chi$`s, in both zigzag and arm-chairs directions are given. As
shown at low temperatures, the exponents for zigzag- and armchairs-
directions are about $0.7$ (see also Ref.~\cite{L}). However, the
exponent for armchairs direction is greater than the one for zigzag
direction at large enough temperatures. We have also used the bond
order potential proposed by Ghiringhelli {\it et al.}~\cite{los},
and found good agreement between the results of the two potentials
~\cite{Faso}. This is physically understandable, since our
simulation has been performed for $T < 700K$ and the two potentials
mainly differ at higher temperatures. Fig.~3 (inset) shows also the
amplitude of the second moments, $C_x$ and $C_y$, which are defined
as, $S_x(r_x) = C_x r_x ^{{\xi_2}_x}$ and $S_y(r_y) = C_y r_y
^{{\xi_2} _y}$ in the scaling region, in terms of temperature.

To determine the characteristic length scales, we define the
quantity $Q(r)$ as the difference between the joint probability
distribution function (PDF) of height fluctuations at two points.
For instance, given $y$ and $y+r$ points, the $Q(r)$ is calculated
by $P(h_1,y;h_2,y+r)$ and product of two PDFs, $P(h_1,y) $ and
$P(h_2,y+r)$~\cite{CMB}. Thus, \be Q(r) = \int dh_1 dh_2 |
P(h_1,y;h_2,y+r) - P(h_1,y) P(h_2,y+r)|~. \ee
 In Fig.~4 we have shown $Q(r)$ as a function of
 $r$. This figure clearly indicates that the height fluctuations
at scales of $90{\AA}$ and $125{\AA}$ for $T=300K$ and $T=30K$,
respectively, are almost independent. Obviously, the $Q(r)$ becomes
$r-$ independent after these values of $r$ (by considering its error
bars). Furthermore, we have found the same value for correlation
length for graphene with fixed boundary condition. As a consequence,
this length scale is not an artifact of boundary conditions.

\section{MARTIN-SIGGIA-ROSE EFFECTIVE ACTION}
Roughness of graphene results in a random gauge field and affects
its electronic structure~\cite{iordanskii}. The dependence of the
hopping integral $\Gamma$ on the deformation tensor is expressed by
$ \Gamma = \Gamma_0 + \frac{\partial \Gamma}{\partial u_{ij}}
u_{ij}$, where $u_{ij}$~\cite{nelson} is given by $ u_{ij} =
\frac{1}{2} \{ \frac{\partial u_i}{\partial x_j} + \frac{\partial
u_j}{\partial x_i} + \frac{\partial u_k}{\partial x_i}
\frac{\partial u_k}{\partial x_j} + \frac{\partial h}{\partial x_i}
\frac{\partial h}{\partial x_j} \}$. Here, $x_i\equiv(x,y)$ are
coordinates in the plane and $u_i$ are the corresponding components
of the displacement vector. In the presence of roughness, an
effective Dirac Hamiltonian describes the electron states near the
K-point, $ H= v_F \sigma ( -i \hbar \nabla - \frac{e}{c} \bf A )$,
where $v_F = \sqrt{3} \Gamma_0 a / 2 \hbar $ and $\bf A$ is the
gauge field. The gauge field can be written in terms of the hopping
integral $\Gamma$ as: $A_x= \frac{c}{2 e v_F} ( \Gamma_2 + \Gamma_3
- 2 \Gamma_1$ ) and $A_y = \frac{\sqrt{3} c}{2 e v_F} ( \Gamma_3 -
\Gamma_2)$~\cite{katsnelson}. Labels $1,2$ and $3$ refer to the
nearest neighbors atoms with vectors $( -a/\sqrt{3}, 0), (
a/2\sqrt{3}, -a/2)$ and $ (a/2\sqrt{3}, a/2)$, respectively. In
Fig.~5, we have plotted the structure function of the gauge field
$\bf A$, namely $S_A (r) = < | (\bf A(\bf x + \bf r) - \bf A ( \bf
x)) \cdot \hat x |^2 > $ versus the scale in $y$-direction. The
vectors $\bf r$ and $\hat {\bf x}$ can be chosen in $x-$ and $y-$
directions. Therefore, we have four different types of structure
functions. As shown in Fig.~5, the gauge field and the related
magnetic field have small scale correlation with correlation length
$l_c \simeq 2-3{\AA}$. We have checked the other three structure
functions of the induced gauge field and their cross correlation
functions and find that they have small scale correlation and are
almost statistically independent.

In what follows we have developed a renormalization group analysis
to investigate the Dirac equation in random gauge field and show
that no delocalization-localization transition occurs for electrons
in such a random gauge potential$^{20,21}$.

The Lagrangian of Dirac fermions in (2+1) dimensions and in the
presence of gauge potential $A_{\mu}$ is given by: $
{\mathcal{L}}=i\int dt \int d^{2}{\bf x}
\bar{\psi}\gamma^{\mu}(\partial_{\mu}-A_{\mu})\psi $, in which
$\gamma^{0}=\sigma_{z},\gamma^{1}=i\sigma_{y},
\gamma^{2}=-i\sigma_{x}$, $\sigma$'s are the Pauli matrices and
$\gamma$'s satisfy the Clifford algebra
$\{\gamma^{\mu},\gamma^{\nu}\}=2g^{\mu \nu}$. The wave functions
$\psi(\bf x,t)$ and $\bar{\psi}(\bf x,t)$ are 2D Dirac spinors and
$A_{\mu}$ is a static random gauge field with a Gaussian
distribution having zero mean value. The covariance is given by $
\langle(A_{i}({\bf x})A_{j}({\bf
x}')\rangle=2D_{0}\delta_{ij}\delta({\bf x}-{\bf x}')$ where $ i,j
\equiv 1,2$ and $D_{0}$ is the intensity of its spatial
fluctuations. This relation shows the spatially uncorrelated nature
of the gauge filed. We note that the Fermi velocity of electrons in
graphene is of the order of $10^6$ m/s and the typical velocity of
the height fluctuations is of the order of $30$ m/s. Therefore, the
random gauge potential will act as a quenched random field on
electrons. Focusing on a single mode with energy $\epsilon$ we get
the following expression for the Lagrangian:

\be
{\label {lagrangian2}} {\mathcal{L}}=\int d^{2}{\bf x}
\bar{\psi}(i\gamma^{k}\partial_{k}-i\gamma^{k}A_{k}+\epsilon\gamma^{0})\psi~.
\ee

The expectation value of any operator $\mathcal{O}$ can be
calculated as follows

\be \langle {\mathcal{O}}\rangle=\frac{\int
{\mathcal{D}}\psi{\mathcal{D}}{\bar\psi} {\mathcal{O}}
\exp(-i{\mathcal{L}})}{Z}~, \ee in which $Z$ the partition function
is defined by
\bea
Z&=&\int {\mathcal{D}}\psi{\mathcal{D}}{\bar\psi}
{\mathcal{O}} \exp(\int d^{2}{\bf x}
\bar{\psi}(-\gamma^{k}\partial_{k}+\gamma^{k}A_{k}-i\epsilon\gamma^{0})\psi
) \nonumber \\
&=&
\det(-\gamma^{k}\partial_{k}+\gamma^{k}A_{k}-i\epsilon\gamma^{0})~.
\eea

Introducing the Dirac bosons $\chi$ and ${\bar\chi}$, one can
re-expressing the above determinant as follows

\be
Z^{-1}=\int {\mathcal{D}}\chi{\mathcal{D}}{\bar\chi}
\exp\left(-\int d^{2}{\bf x}
\bar{\chi}(-\gamma^{k}\partial_{k}+\gamma^{k}A_{k}-i\epsilon\gamma^{0})\chi\right),
\ee

and implicitly we have

\be \langle {\mathcal{O}}\rangle=\int
{\mathcal{D}}\psi{\mathcal{D}}{\bar\psi}
{\mathcal{D}}\chi{\mathcal{D}}{\bar\chi}{\mathcal{O}} \exp\{\int
d^{2}{\bf x}
\bar{\psi}(-\gamma^{k}\partial_{k}+\gamma^{k}A_{k}-i\epsilon\gamma^{0})\psi
-\int d^{2}{\bf x}
\bar{\chi}(-i\gamma^{k}\partial_{k}+\gamma^{k}A_{k}-i\epsilon\gamma^{0})\chi\}.
\ee

Now by integrating the above result over the Gaussian variable
$A_{\mu}$ whose probability density function is given by

\be P\propto \exp(-\frac{1}{4D_{0}}\int d{\bf x} A_{k}^{2}), \ee one
reaches the following result for the averaging of expectation values
over quenched random gauge
 \be \langle\langle {\mathcal{O}}\rangle\rangle=\int
{\mathcal{D}}\psi{\mathcal{D}}{\bar\psi}
{\mathcal{D}}\chi{\mathcal{D}}{\bar\chi}
{\mathcal{O}}\exp(-S_{0}-S_{int})~, \ee where the free part of the
effective action is

\be S_{0}=\int d^{2}{\bf x}
\bar{\psi}(\gamma^{k}\partial_{k}+i\epsilon\gamma^{0})\psi +\int
d^{2}{\bf x}
\bar{\chi}(\gamma^{k}\partial_{k}+i\epsilon\gamma^{0})\chi~, \ee and
the interaction part

 \be S_{int}=-D_{0}\int
d^{2}{\bf
x}\left[\left({\bar\psi}\gamma^{k}\psi+{\bar\chi}\gamma^{k}\chi)\right)\cdot
\left({\bar\psi}\gamma^{k}\psi+{\bar\chi}\gamma^{k}\chi\right)\right]~,
\ee where $k=1,2$. The $\beta$ function of the coupling $D_0$ will
determine its behavior under changing the scale.
 We found that in one-loop order the correction to the short-range roughness
 intensity ($D_0$) is proportional to $k^2$ , therefore this correction vanishes
 in long-wave length limit. One can see that the same will happen at higher order
 of of perturbation, leading us to the conclusion that the vanishing of the beta function in all orders should be
 the consequence of the Ward  identity due the conservation of
 Dirac current which leads to an incompressible flow of electrons in the low energy limit( $\epsilon \sim 0$).

 This result shows that the
resistance against the electron flow due the interaction of Dirac
fermions and roughness of the Graphene, remains unchanged under
renormalization group flow towards large scales, and this in
turn, excludes the possibility of the localization of low energy
states (For more details, see for instance
Ref[\onlinecite{morpurgo,morozov,demartino,aleiner, ziegler}]).

\section{Summary}

In Conclusion, we find the temperature dependence of the roughness
exponents in different directions of a graphene sheet by simulating
the surface within molecular dynamics approach. We have used the
Brenner empirical inter-atomic interactions for graphite which is a
semiconductor. The point
we are making in this publication is the further detailed explanations
raised in Ref.[\onlinecite{Faso}] which
roughness could affect. We answer the question whether the roughness
could lead to localized electrons in graphene or not?. The
correlation function of height fluctuations shows that depending on
the temperature, there are characteristic length scales in the order
of $ \approx 90 {\AA}$ at room temperature. We show that the induced
gauge field has a short- range nature with correlation lengths
approximately $\simeq 2-3 {\AA}$. More importantly, roughness
essentially can affect to the electronic properties like
conductivity and modulation of the hopping integrals~\cite{castro}.
We treat the problem analytically by using the Martin-Siggia-Rose
method. The renormalization group flows do not yield any
delocalized-localized transition due to roughness. In the present
work, the effect of Dirac-like electrons on roughness are not
considered. It would be of intrest to develop our work may use the
quantum molecular dynamics simulation or ab initio Car-Parinnello
molecular dynamics for Dirac-like electrons to investigate the
dependence of Dirac-like electron on the graphene roughness.
\begin{acknowledgments}
We are grateful to A. Geim for illuminating discussions
and comments. M.R.R.T. would like to express his deep
gratitude to the Alexander von Humboldt Foundation and
Universitat Oldenburg for their financial support and providing
an excellent environment for research.
\end{acknowledgments}

\newpage

\begin{figure}[t]
\begin{center}
\includegraphics[width=1.\linewidth]{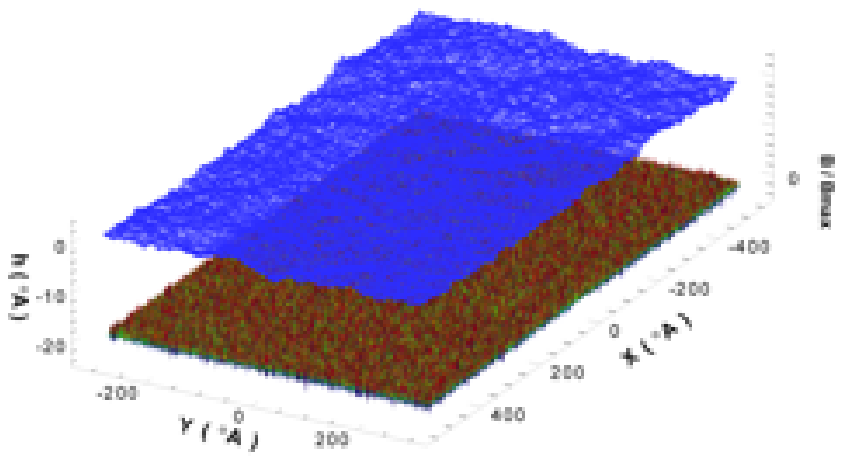}
\caption{(color online) Snapshot of the graphene surface at $300 K$
(upper graph). The sample contains a lattice size of $500 \times
500$ atoms. In the lower graph, we have plotted the induced magnetic
field due to the roughness of graphene surface.}
\end{center}
\end{figure}

\newpage


\begin{figure}[t]
\begin{center}
\includegraphics[width=0.8\linewidth]{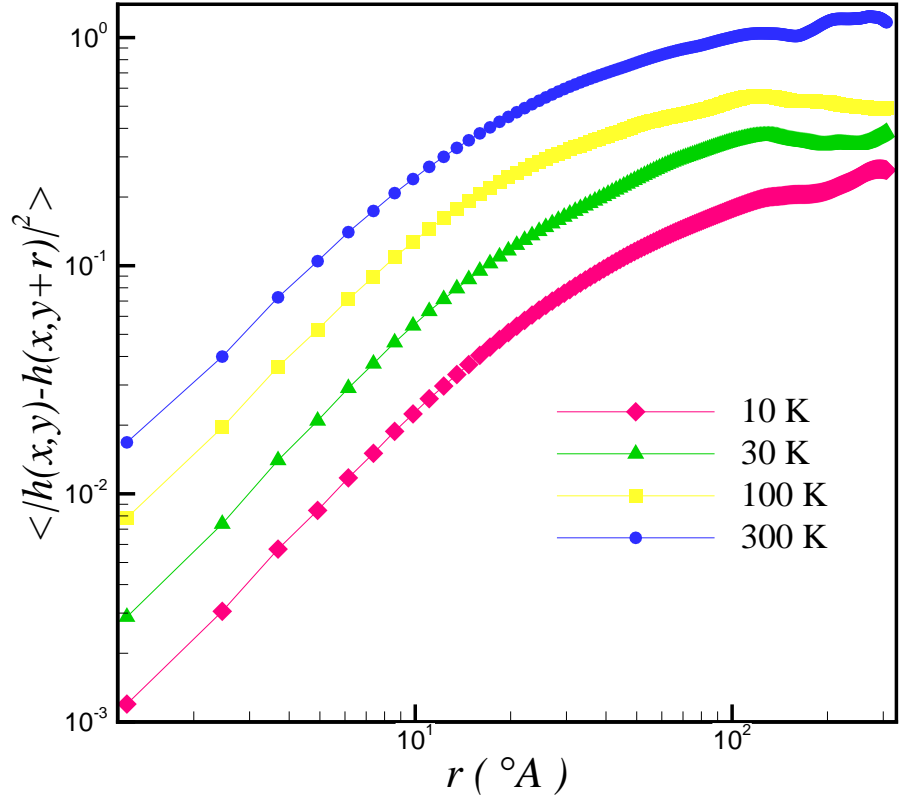}
\caption{(color online) Log-log plot of the second moment of height
difference as a function of $r$, in $y$ (arm-chairs) direction,
which shows that for samples with temperatures $10,30,100$ and $300$
$K$ the hight fluctuations have scaling behavior in small scales. It
indicates that the roughness exponent (the slope of the plots)
decreases with temperature, and means that the surface will be rough
at high temperatures. A similar figure can be found for the
height-height structure function in $x $ (zigzag) direction. }
\end{center}
\end{figure}

\begin{figure}[t]
\begin{center}
\includegraphics[width=0.8\linewidth]{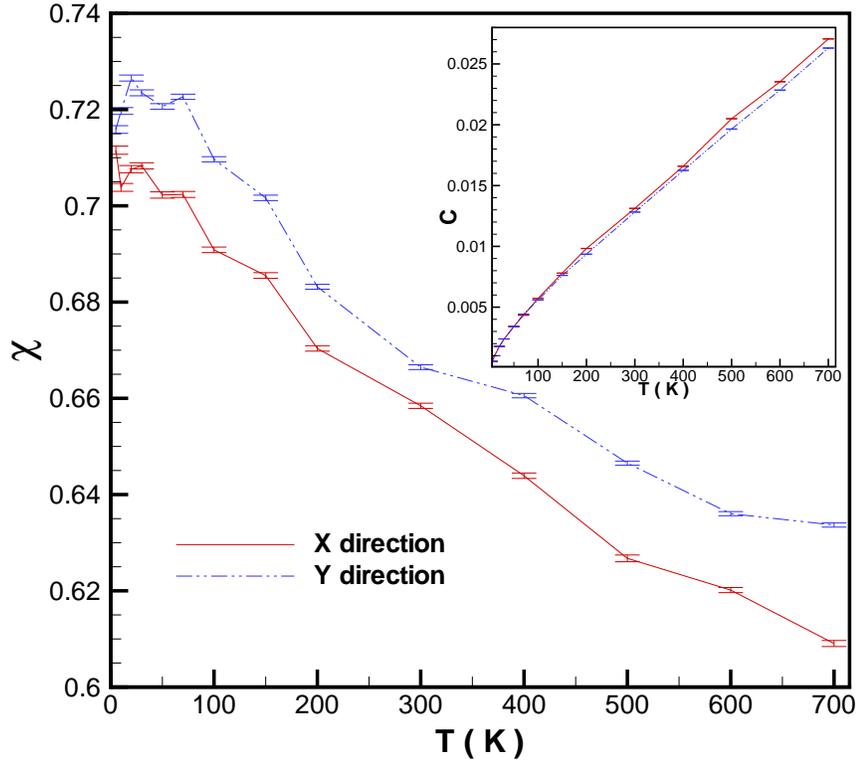}
\caption{(color online) The temperature dependence of the roughness
exponents show that graphene is smoother in arm-chairs direction as
compared to zigzag direction. For very smooth surface the exponent
will be $\sim 1.0$. At high temperature limit the exponents approach
 a random noise exponent (i.e. 0.5). The anisotropy of the graphene
is due to
 the fact that the lattice spacing in arm-chairs and zigzag directions are
different. It shows  that (inset) the amplitude of the second
moments, $C_x$ and $C_y$, are increasing function of temperature. }
\end{center}
\end{figure}


\newpage

\begin{figure}[t]
\begin{center}
\includegraphics[width=0.8\linewidth]{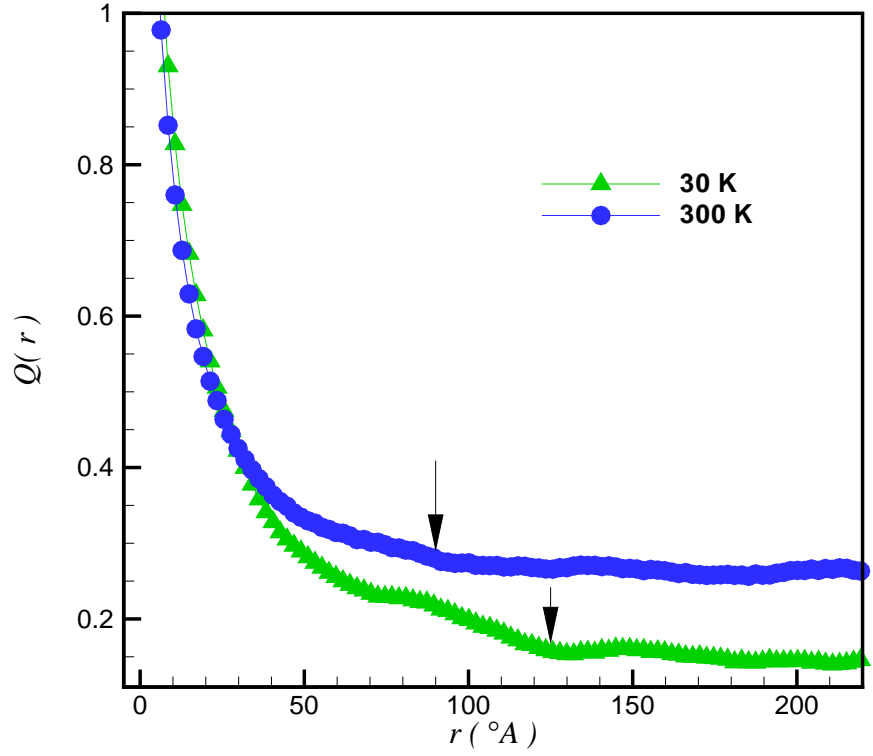}
\caption{(color online)Scale dependence of $Q(r)$, defined by Eq.(1)
as a function of $r$. It shows that the height fluctuations have
characteristic scales of $90{\AA}$ and $ 125{\AA}$ for $T=300K$ and
$T=30K$, respectively. }
\end{center}
\end{figure}


\begin{figure}[t]
\begin{center}
\includegraphics[width=0.8\linewidth]{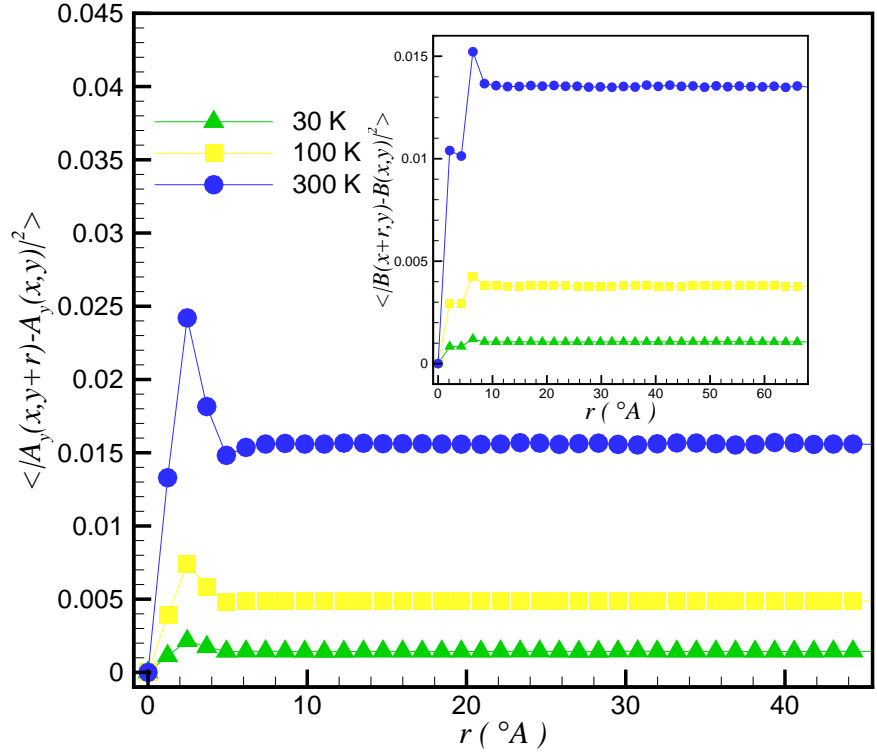}
\caption{(color online) Log-log plot of the second moment of the
induced gauge field (inset structure function of magnetic field) in
$y$ direction, for samples with temperatures $30,100$ and $300$ $K$.
The correlation length is about  $2-3 {\AA}$.  Similar figures can
be found for the other three gauge field structure functions.}
\end{center}
\end{figure}

\end{document}